\documentclass[draftclsnofoot, onecolumn]{IEEEtran}
\usepackage[latin1]{inputenc}
\usepackage{amssymb}
\usepackage[dvips,pdftex]{graphicx}
\usepackage{amssymb}
\usepackage{caption}
\usepackage{subcaption}

\usepackage{verbatim}
\usepackage[normalem]{ulem}
\usepackage{lineno}
\usepackage{setspace}
\usepackage{color}
\usepackage{url}
\usepackage{cite}

\usepackage{float}
\usepackage{xspace}
\usepackage{multirow}

\usepackage[stable]{footmisc}

\usepackage{amsmath}

\IEEEoverridecommandlockouts

\renewcommand{\textbf}[1]{\begingroup\bfseries\mathversion{bold}#1\endgroup}

\begin{document}

\title{From Co- Toward Multi-Simulation of Smart Grids based on HLA and FMI Standards: A Telecontrol Case Study Based on Real World Configurations
\thanks{This work was supported by NSERC Strategic Project Grant No. 413427-2011, FQRNT Doctoral Research Scholarship No. 165516, and SYTACom Internship Program. The work is part of the Information System And Telecom Integrated Simulation for Smart grids (ISATISS) project at EDF R\&D.}
}


\author{Martin L\'evesque and Martin Maier, Optical Zeitgeist Laboratory, INRS, Canada\\Christophe B\'echet, Eric Suignard, and Anne Picault, EDF R\&D, Clamart, France\\G\'eza Jo\'os, Department of Electrical and Computer Engineering, McGill University, Canada

\thanks{Corresponding author: Martin L\'evesque, INRS, Montr\'eal, QC, Canada H5A 1K6 (email: \texttt{levesquem@emt.inrs.ca}).}
}


\IEEEaftertitletext{\vspace*{-.7cm}
}

\maketitle
\begin{abstract}
In this paper, a multi-simulation model is proposed to measure the performance of all Smart Grid perspectives as defined in the IEEE P2030 standard. As a preliminary implementation, a novel information technology (IT) and communication multi-simulator is developed following an High Level Architecture (HLA). To illustrate the usefulness of such a multi-simulator, a case study of a distribution network operation application is presented using real-world topology configurations with realistic communication traffic based on IEC 61850. The multi-simulator allows to quantify, in terms of communication delay and system reliability, the impacts of aggregating all traffic on a low-capacity wireless link based on Digital Mobile Radio (DMR) when a Long Term Evolution (LTE) network failure occurs. The case study illustrates that such a multi-simulator can be used to experiment new smart grid mechanisms and verify their impacts on all smart grid perspectives in an automated manner. Even more importantly, multi-simulation can prevent problems before modifying/upgrading a smart grid and thus potentially reduce costs to the utility.
\end{abstract}


\section{Introduction}

\PARstart{T}{he IEEE} P2030 standard is one of the first attempts to standardize the smart grid \cite{standardP2030}. It decomposes a smart grid into three fundamental perspectives:

\begin{itemize}
\item \textit{Power systems}: This perspective deals with the generation, delivery, and consumption of electrical energy.
\item \textit{Communication technology}: It defines the integration of networking components and communication protocols. 
\item \textit{Information technology (IT)}: The IT perspective processes and controls the data flow related to applications, which operate and manage the power systems.
\end{itemize}

	Each of these three perspectives are widely studied using sophisticated continuous, event-driven, and flow-based simulators. In order to integrate simulators, standards have been developed recently, such as the Functional Mockup Interface (FMI) and IEEE 1516 High Level Architecture (HLA). The FMI standard is the best suited for simulators with differential, algebraic, or discrete equations, whereas HLA fits best to any type of distributed event-based simulators. 
	
	The co-simulation of the communication and power system perspectives has recently attracted attention in the smart grid research community \cite{geco, epochs, intVGR, coSimSurvey}. These recent works are clearly a step further toward the multidisciplinary study of smart grids. However, IT has not been considered as a distinct perspective. In reality, the IT systems of a smart grid are quite complex and cannot be, for example, implemented as an application in a network simulator. The business models, operations between agents, and systems, could be modeled with a proper tool such as Enterprise Architect\footnote{Please refer to \url{http://www.sparxsystems.com/products/ea/} for further information about Enterprise Architect.} using flow charts to define complex interactions between IT systems. The multi-agent simulator developed in \cite{multiAgent2} (MASGriP) could also be used, as it allows the simulation of technical and economical activities of several players. Toward this end, in this paper, the IT perspective is added in order to define a multi-simulator modeling all smart grid perspectives. A co-simulation represents a special multi-simulation case characterized by the use of two simulators, whereas multi-simulation is a more generic term for the execution of multiple simulators interacting with each other. For the remainder of the paper, the terms co-simulation and multi-simulation are used for referring to the use of only two simulators and two or more simulators, respectively. Multi-simulation is part of the research vision at \'Electricit\'e de France (EDF) R\&D, the network laboratories devoted to smart grids of the world's largest producer of electricity, EDF, and is planned to be used as a way to validate smart grids and take better engineering decisions. As a preliminary implementation and investigation, \emph{a novel IT and communication smart grid multi-simulator is developed in this work}. Since, to the best of our knowledge, no previous work studied the co-simulation of the IT and communication perspectives, the focus of this paper is on these two specific perspectives. In order to show the usefulness of such a multi-simulator, a case study of telecontrol and monitoring of distribution grids and distributed energy resources (DERs) is conducted. The topology is based on real-world configurations in France for a given distribution grid is modeled and standardized IEC 61850 messages are injected with realistic data rates in order to have reasonable conditions in the communication perspective, which is modeled using Long Term Evolution (LTE) and dedicated Digital Mobile Radio (DMR) wireless networks. LTE is a promising communication technology for smart grids. In \cite{LTE1}, the authors have shown that LTE can satisfy the latency and reliability requirements of distribution automation (DA) networks. Furthermore, in \cite{LTE2}, a worst-case usage factor was considered using LTE for smart grid communication, where all smart grid meters simultaneously detect and report a failure. By using simple yet efficient solutions to alleviate this problem it was shown that LTE be capable of satisfying the quality-of-service (QoS) criteria for smart grid traffic without having a detrimental impact on LTE traffic. In this paper, we investigate a scenario, where the LTE network becomes unavailable. The performance of LTE traffic re-routing towards a low-capacity DMR link is evaluated from both the IT and communication perspectives. We investigate this case study since given that public LTE networks are typically shared with other applications/clients the utility does not have full control over the network, as opposed to a dedicated link such as DMR. A rate adaptation QoS mechanism is also proposed in order to forward all traffic over the low-capacity DMR link when the LTE network fails and compares its performance to a well-known QoS mechanism. This case study is developed to illustrate that smart grid multi-simulators can be used by engineers as a powerful research tool to test new smart grid algorithms \footnote{By smart grid algorithms, we refer to algorithms applied to power systems that take advantage of advanced communication infrastructures, e.g., electric vehicle coordination algorithms.} or mechanisms and validate their impacts on all smart grid perspectives.

	The remainder of this paper is structured as follows. A survey on the key co-simulation concepts and related work is given in Section \ref{sec:stateOfTheArt}. Section \ref{sec:multiSimFramework} describes the proposed multi-simulation model for smart grids. The context of the case study is defined in Section \ref{sec:caseStudy}. Numerical results are presented and discussed in Section \ref{sec:numericalResults}. Finally, conclusions are drawn in Section \ref{sec:conclusions}.	

\section{Communication and Power System Co-Simulation}
\label{sec:commPowerCoSim}

A smart grid contains multiple different perspectives, as described in the previous section. To model each of these perspectives, one may either implement all perspectives in a given simulation environment or couple available simulators. Recently, Mets \emph{et al.} conducted a comprehensive survey on smart grid simulations focusing on combined power and communication network simulation \cite{coSimSurvey}. They concluded and observed that:

\begin{itemize}
\item There are two main types of studies: ($i$) wide-area monitoring, protection, and control, and ($ii$) demand response. In Section \ref{sec:caseStudy}, we will perform a case study of type ($i$) for a smart telecontrol application based on real-world configurations.
\item Combined communication and power system simulations can be realized by the use of co-simulation or integrated simulation. Integrated simulation is typically used when one of both perspectives (either power or communication) can be significantly abstracted. On the other hand, when a more detailed simulation is required, the co-simulation approach is preferred by reusing existing tools.
\item Combined simulation is a challenging task since it needs to synchronize operations and states, especially time, which is discussed in greater detail in the next section.
\item Federation-based smart grid simulation, which is the focus of this work, is a promising approach to allow large-scale smart grid simulations. The authors of \cite{coSimSurvey} concluded that the use of standards, such as the one discussed in the next section, will play a key role in combining several existing simulators.
\end{itemize}

\begin{figure}
\centering
\includegraphics[width=.70\textwidth]{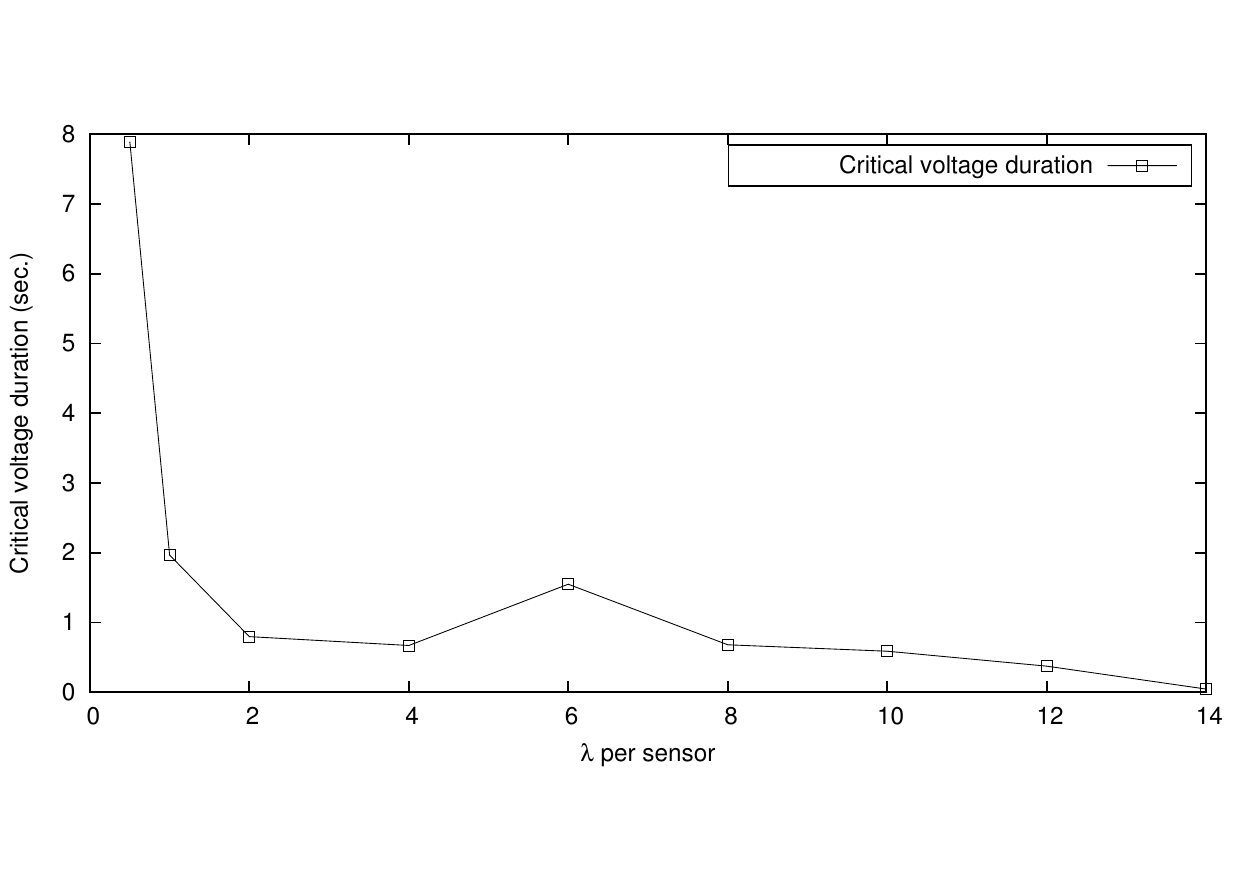}
\caption{Performance evaluation of an electric vehicle coordination algorithm by means of co-simulation - Critical voltage duration (power system perspective) as a function of the data rate of sensors (communication perspective)  \cite{coSim4}.}
\label{fig:cosimResult1}
\end{figure}

\begin{figure}
\centering
\includegraphics[width=.70\textwidth]{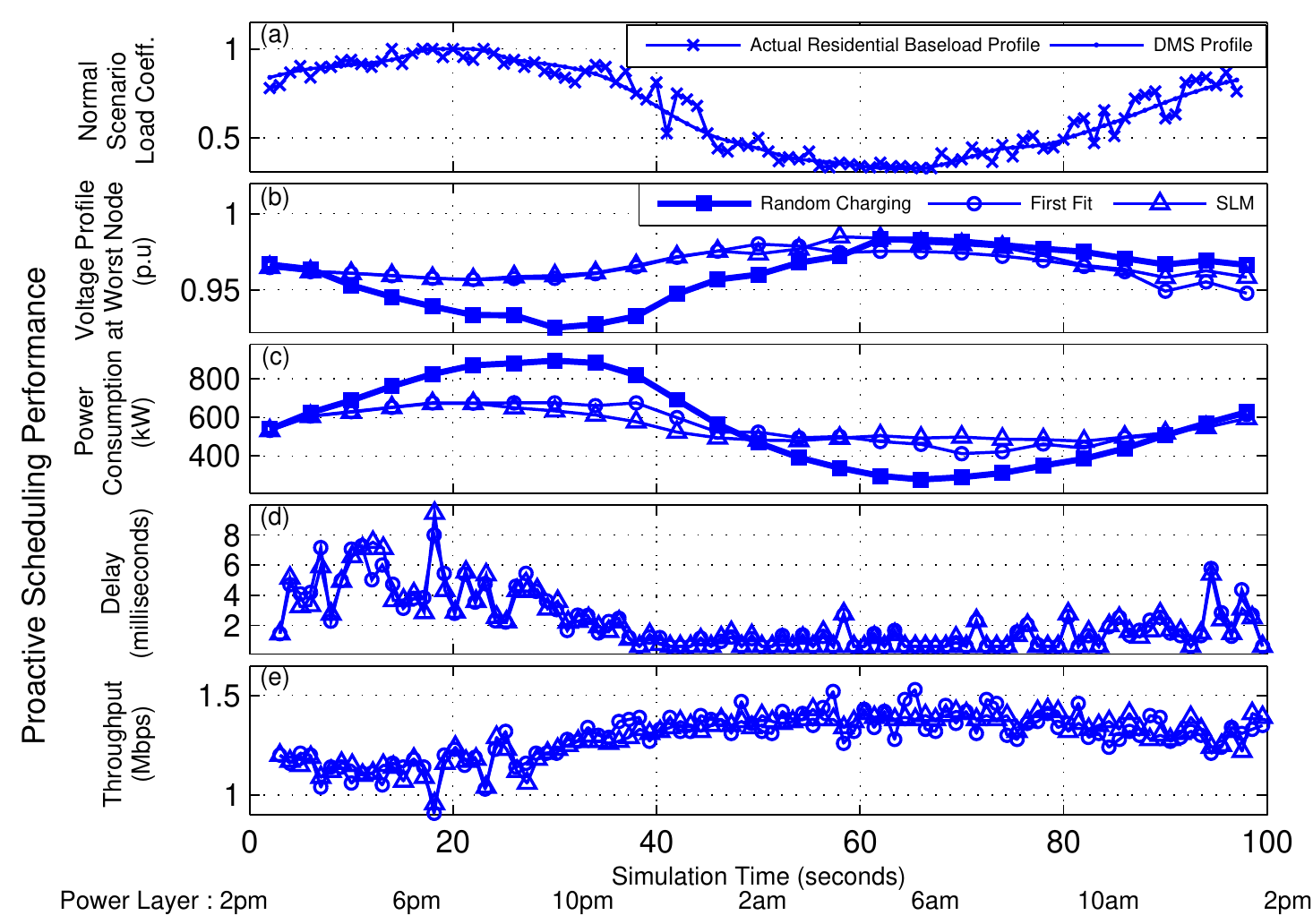}
\caption{Performance evaluation of different electric vehicle charging strategies by means of co-simulation - Power system perspective: (a) load profile, (b) voltage, and (c) power consumption; Communication perspective: (d) end-to-end delay and (e) throughput \cite{coSim5}.}
\label{fig:cosimResult2}
\end{figure}

	Furthermore, the authors concluded that use cases with a focus on demand response generally adopt multi-agent systems. The current trends to control and monitor the operation of electric power systems are moving toward the use of automated agent technology, known as multi-agent systems \cite{multiAgent1}. The main contributions achieved by modeling both the communication and power system perspectives are as follows:

\begin{itemize}
\item The operation of special protection schemes (SPS), designed to counteract power system instability, were shown to be highly affected by the communication loss rate using the electric power and communication synchronizing simulator (EPOCHS) \cite{epochs}.
\item By means of co-simulation, the authors of \cite{coSim4} have shown that for a given electric vehicle control scheme the critical voltage fluctuations are highly influenced by the data rate of sensors, as depicted in Fig. \ref{fig:cosimResult1}.
\item In \cite{geco}, a power sytem protection scheme was validated to operate within the required threshold of 100 ms under different scenarios, whereby the communication infrastructure is the main delay component. Deng \emph{et al.} also verified by means of co-simulation several protection applications having 50-100 ms time constraints \cite{WAMPaper}.
\item In \cite{coSim5}, different electric vehicle charging strategies were evaluated by means of co-simulation using real-time exchange of messages, as depicted in Fig. \ref{fig:cosimResult2}. The results report on different metrics from both the power system and communication perspectives for each time of the day.
\item A proposed integrated vehicle-to-grid, grid-to-vehicle, and renewable energy sources (IntVGR) coordination scheme has been co-simulated over a converged fiber-wireless broadband access network \cite{intVGR}. The co-simulator allowed to quantify the throughput and delay of using such a control scheme for different times of day and yearly seasons, as well as show the improvement from a power system perspective. 
\end{itemize}

\section{Time Synchronization and Multi-Simulation Standards}
\label{sec:stateOfTheArt}

	As concluded in \cite{coSimSurvey}, when detailed simulations are required for integrated communication and power systems, the co-simulation approach is preferred using existing tools and standards. Although co-simulation frameworks allow to evaluate novel smart grid control schemes, quantify the impact of the communication perspective on the power system, combining simulators is not necessarily trivial. One main requirement of a co-simulator is time synchronization, which involves a number of critical issues. In this section, we review models and standards used to integrate multiple simulators to form integrated multi-simulations.

\subsection{Time Synchronization}

	Each simulator can have a simulation time frame and progresses according to a scheduler managing events and/or functions. The synchronization mechanism acts as mediator in order to make both simulators progress simultaneously. There exist 3 main synchronization approaches \cite{geco}:
	
\begin{itemize}	
\item \textit{Time-stepped:} In \cite{HLAFirstPaper}, the paper that first presented the HLA architecture, the authors proposed to discretize the multi-simulation into slots with duration $\tau$. At the beginning of each timeslot, simulators notify their messages during the upcoming timeslot. Then for each given message, subscribed simulators are stopped at the message time arrival. However, if new messages are created during a given timeslot, these new messages will notify other simulators only at the end of the current timeslot, as illustred in Section 3 of \cite{epochs}. To mitigate this issue, $\tau$ should be set sufficiently small. The time-stepped approach has been widely used in \cite{fmiRTI, smartGridComAndCoSim, interfacingIssues, hlaOmnet, hla2, epochs, coSimDistrControls, coSim}.
\item \textit{Global event-driven:} In this approach, a single simulation time frame is being used by both simulators \cite{geco, coSim2}.
\item \textit{Without synchronization:} In some co-simulation models, only one simulation time frame is used, since one simulator is time independent. One example is the power flow analysis calculation with OpenDSS driven by a communication simulator (ns-2/3, OMNeT++, etc.) \cite{coSim3, coSim4, intVGR}.
\end{itemize}

	The global event-driven approach has the advantage of avoiding all potential synchronization errors. However, in the context of multi-simulation with the integration of multiple continuous and event-driven simulators, using this approach causes a given simulator to completely depend on another simulator. As simulators can change and evolve, each simulator should be independent of other simulators for smooth upgrades. The time-stepped approach allows each simulator to act as an independent entity, which can interact with one or several simulators for multidisciplinary studies. Thus, this approach simplifies the re-utilization of existing off-the-shelf simulators. Furthermore, this approach, which is considered in this work, is being adopted by the HLA standard, described in the following.
	
	Synchronization and simulation coupling was the topic of many simulation studies. In \cite{sync2}, the authors proposed several relaxed synchronization techniques aiming at reducing the number of exchanges between federates. The concept of redundant host execution was introduced in \cite{sync3} to minimize the simulation idle time caused by data dependency and to improve parallelism. In \cite{sync4}, a solution for loose coupling of heterogeneous simulation components was proposed, whereby a lightweight message allowed multiple simulators to exchange the same messages. 

\subsection{Multi-Simulation Coupling Standards and Models}
\label{sec:couplingStandards}

	The modeling and simulation community created different standards in order to combine simulators to maintain interoperability.

\subsubsection{IEEE 1516 - High Level Architecture (HLA)}

	The HLA standard defines a simulation interface specification and an run time infrastructure (RTI) allowing to run a set of independent simulators to coordinate them all, thus forming a federation \cite{HLAFirstPaper, hla1}. Each simulator, acting as a \textit{federate}, implements an interface allowing the synchronization by communicating with the RTI. Each federate sends time advance messages to the RTI which sends back grant messages such that all simulators progress in a coordinated manner. A given federation of simulations can be linked to another one by connecting them via bridges. This model, though simple, is versatile and scalable. The HLA framework provides conservative and optimistic synchronization approaches. The conservative approach is the most widely used and relies on a lookahead function to obtain events in the upcoming timestep. Whereas the optimistic approach introduces a rollback method to go back to the last timestep, whenever causality errors occur \cite{sync1}. 

\subsubsection{Functional Mockup Interface (FMI)}

	The functional mockup interface (FMI) standard has been initiated by Daimler to improve the exchange of simulation models in the automotive industry \cite{fmi1}. Simulators supporting FMI provide an interface library referred to as functional mockup units (FMUs). An FMU contains configuration files and C-functions. An FMU can be either imported by another simulator or used to form a co-simulation. However, one main drawback of FMI is that it does not provide any master algorithm to coordinate a set of simulations \cite{fmiRTI}. In \cite{coSimFMI}, Palensky \emph{et al.} developed a novel FMI compliant co-simulation platform based on GridLAB-D and OpenModelica simulators to model electric vehicles, batteries, and distribution grids.

\subsubsection{Combining both HLA and FMI Standards}


	Since some simulators support HLA (e.g., OPNET), others support FMI (e.g., EMTP-RV, Matlab/Simulink), and several do not support either one, it might become difficult to create multi-simulators. In \cite{fmiRTI}, the authors proposed to use the RTI of the HLA standard as a master to FMI components. In fact, both standards should be viewed as complementary, whereby simulators supporting FMI can communicate with other simulators of the federation via HLA.

\section{Multi-Simulation smart grid Model}
\label{sec:multiSimFramework}

\begin{figure}
\centering
\includegraphics[width=.70\textwidth]{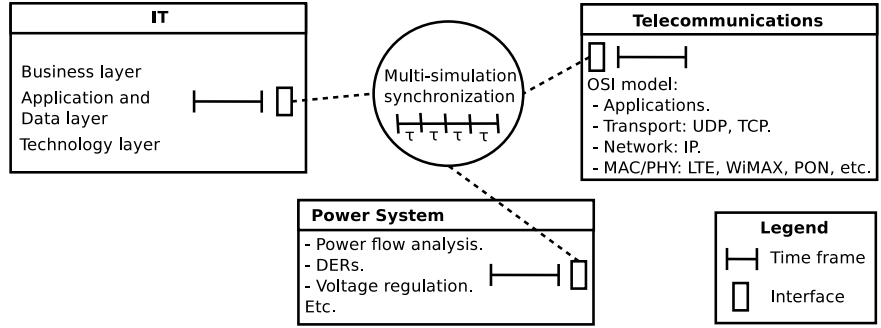}
\caption{Multi-simulation model covering all smart grid perspectives. Each simulator has timing services modeling a time frame and interacts externally by providing interface services.}
\label{fig:multiSimModel}
\end{figure}

\subsection{A Complete Smart Grid Simulation Model}

	As summarized in the previous sections, the co-simulation of the communication and power system perspectives was the subject of several recent novel studies. To simulate and validate a complete smart grid, the simulation model should also take into account the operations related to the business model, use cases among different smart grid entities, processes, and so forth. These high-level interactions control the power system using the communication infrastructure and thus their role is quite significant in the overall operation of the power grid. In \cite{multiAgent2}, the authors proposed MASGriP, a simulation platform to study multi-agent systems for smart grids. Their platform could also be viewed as an IT simulator. The simulation of cyber-attacks was performed in\mbox{\cite{cyberAttackSim}} to expose potential attacks and vulnerabilities of the power grid state estimator. Cyber-security should be part of a complete IT simulator, as described in the following in more detail. As standardized in the IEEE P2030 standard, a smart grid is modeled with the IT, communication, and power system perspectives. Toward this end, our long-term goal is to create a multi-simulator that models all smart grid perspectives, as depicted in Fig. \ref{fig:multiSimModel}. In the proposed model, the IT perspective is decomposed into several layers following the ArchiMate standard \footnote{The ArchiMate specification was created by the Open Group and is available at \url{https://www2.opengroup.org/ogsys/catalog/c091}.}:

\begin{itemize}
\item \textit{Business layer:} This layer defines the business processes and actors.
\item \textit{Application and Data layer:} It models the application services between systems (e.g., Common Information Model (CIM), IEC 61850 messages, etc.). Cyber-security, including the modeling of cyber-attacks, is also part of this layer.
\item \textit{Technology layer:} This layer deals with the communication and hardware infrastructure supporting the Applications and Data layer. 
\end{itemize}	
	
	 
	 The communication perspective is composed of the Open Systems Interconnection (OSI) model layers, from the application layer to the physical layer. The power system perspective models the electrical grid using one or several existing simulators. All simulators are synchronized by a central RTI, whereby the RTI and simulators exchange messages through the use of federates, acting as interfaces. It is clear that there is no need to reinvent the wheel. Off-the-shelf simulators can be used. To do so, multi-simulation standards such as HLA and FMI will play a key role in integrating simulators.
	
	Such a complete multi-simulator should be viewed as a tool for smart grid engineers in order to:
	
\begin{itemize}
\item Quantify the overall smart grid performance when a modification/upgrade is planned. With such a complex system, it will become quite a challenge to know exactly what are the impacts on all smart grid perspectives when a given modification is orchestrated in a given perspective. The multi-simulator can be, for instance, used to predict problems before deployment, which can significantly reduce long-term costs. 
\item For each given perspective, define a set of automated validation tests. This allows engineers and researchers, while creating new smart grid algorithms/mechanisms, to verify whether or not they keep the smart grid under stable conditions, and to what extent.
\end{itemize}

\begin{figure}
\centering
\includegraphics[width=.70\textwidth]{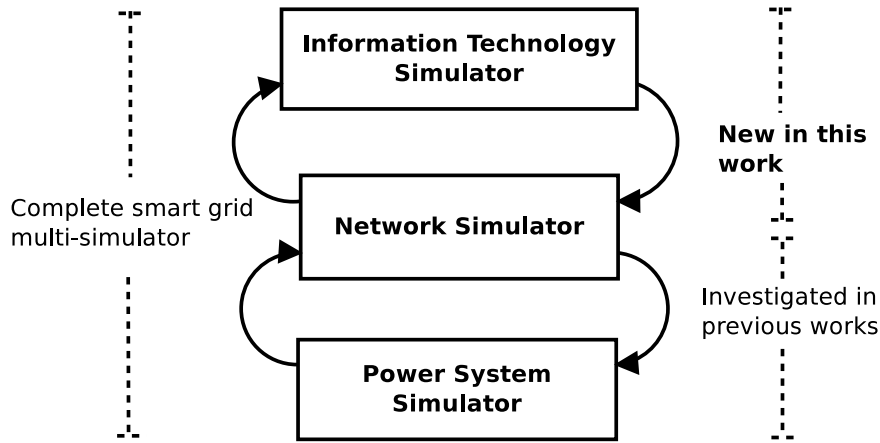}
\caption{Progress on smart grid simulation. Communication power system was previously investigated in \cite{coSimSurvey, epochs, coSim, coSim2, coSim3, coSim4, geco, WAMPaper, coSim5, intVGR}. }
\label{fig:multiSimLayers}
\end{figure}

Fig. \ref{fig:multiSimLayers} depicts the progress on smart grid simulation, whereby communication and power systems were investigated in several recent works. In this paper, the IT and communication perspectives are co-simulated, whereby the next step will be too integrate all three perspectives for complete smart grid simulations. Note that the IT perspective does not communicate directly with the power system, and vice versa, but instead exchanges information via the network simulator. The key towards integration of all three perspectives is the use of RTI to control both IT and communication perspectives as well as the mutual communication and power system relationships.

\subsection{Preliminary Implementation}

\begin{figure}
\centering
\includegraphics[width=.70\textwidth]{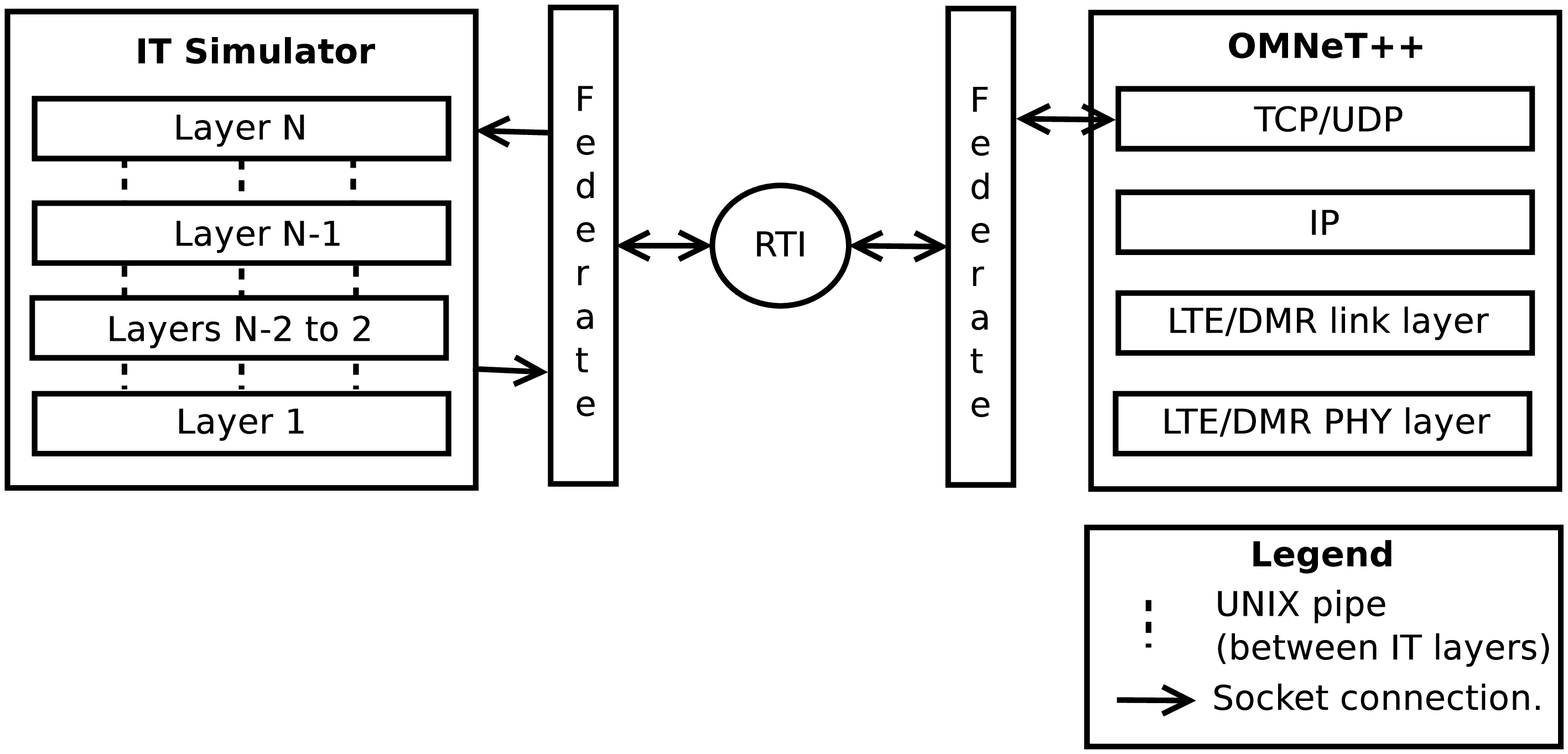}
\caption{Main components of the preliminary multi-simulator implementation.}
\label{fig:preliminaryImplentation}
\end{figure}

	In this paper, the IT and communication perspectives are studied following the HLA standard, since several previous studies have already focused on the communication and power system perspectives. Fig. \ref{fig:preliminaryImplentation} depicts our novel IT and communication multi-simulator. Note that the IT perspective could be modeled in the application layer of the communication simulator. However, this would require to re-implement IT models that already exist in IT simulators, such as Enterprise Architect. Furthermore, such IT simulators contain visual interfaces which can be ran during multi-simulation experiments and IT designers can be used without reinventing the wheels to create rich information flow scenarios.

	The IT perspective is modelled as a set of Java programs generating IEC 61850 messages between nodes of the topology, forming a multi-layer system. All layers receive GRANT messages from the IT federate, which allow the IT simulator to progress to a given GRANT time. Each IT layer forwards its messages to the lower and upper layers, whereby a given layer can add information/header fields to the arriving messages. When the IT simulation time equals the GRANT time, messages are sent to the IT federate, which in turn forwards messages to the RTI, OMNeT++ federate, and finally OMNeT++. 

	For the communication perspective, the HLA-OMNeT++ simulation model \cite{hlaOmnet} is extended and components are added for the LTE and DMR technologies. The RTI component manages time for both simulators by discretizing the time frame into slots with duration $\tau$. The parameter $\tau$ is one of the most important parameters in the proposed multi-simulator and is extensively investigated in Section \ref{subsec:timeslot}.

	A minimal RTI implementation has been realized to conduct research on synchronization algorithms since the commercial RTI framework synchronization algorithms cannot be modified/extended. We found that adding synchronization algorithms in the open source would require significant implementation effort. The time synchronization algorithm developed is defined as follows. Each federate progresses its time frame on a per-timeslot $\tau$ basis coordinated by the RTI, which in turn sends GRANT messages. Messages arriving during a given timeslot are queued. Then, at each synchronization point, queued messages are notified to other federates. Therefore, to reduce synchronization errors, $\tau$ is set sufficiently small such that the time interval $\tau$ becomes negligible for all simulators (e.g., $10^{-5}$ for simulators working at ms time sampling). Compared to the time synchronization algorithm  defined in \cite{HLAFirstPaper}, this algorithm has the inconvenience of introducing a possible delay of maximally $\tau$, but it decreases the number of interactions between the federates and RTI, since interactions occur only at each synchronization point. This minimal RTI could be replaced by an existing HLA library such as Portico.

\subsection{Ongoing and Future Research}
\label{sec:ongoingAndFutureResearch}
\begin{figure}
\centering
\includegraphics[width=.70\textwidth]{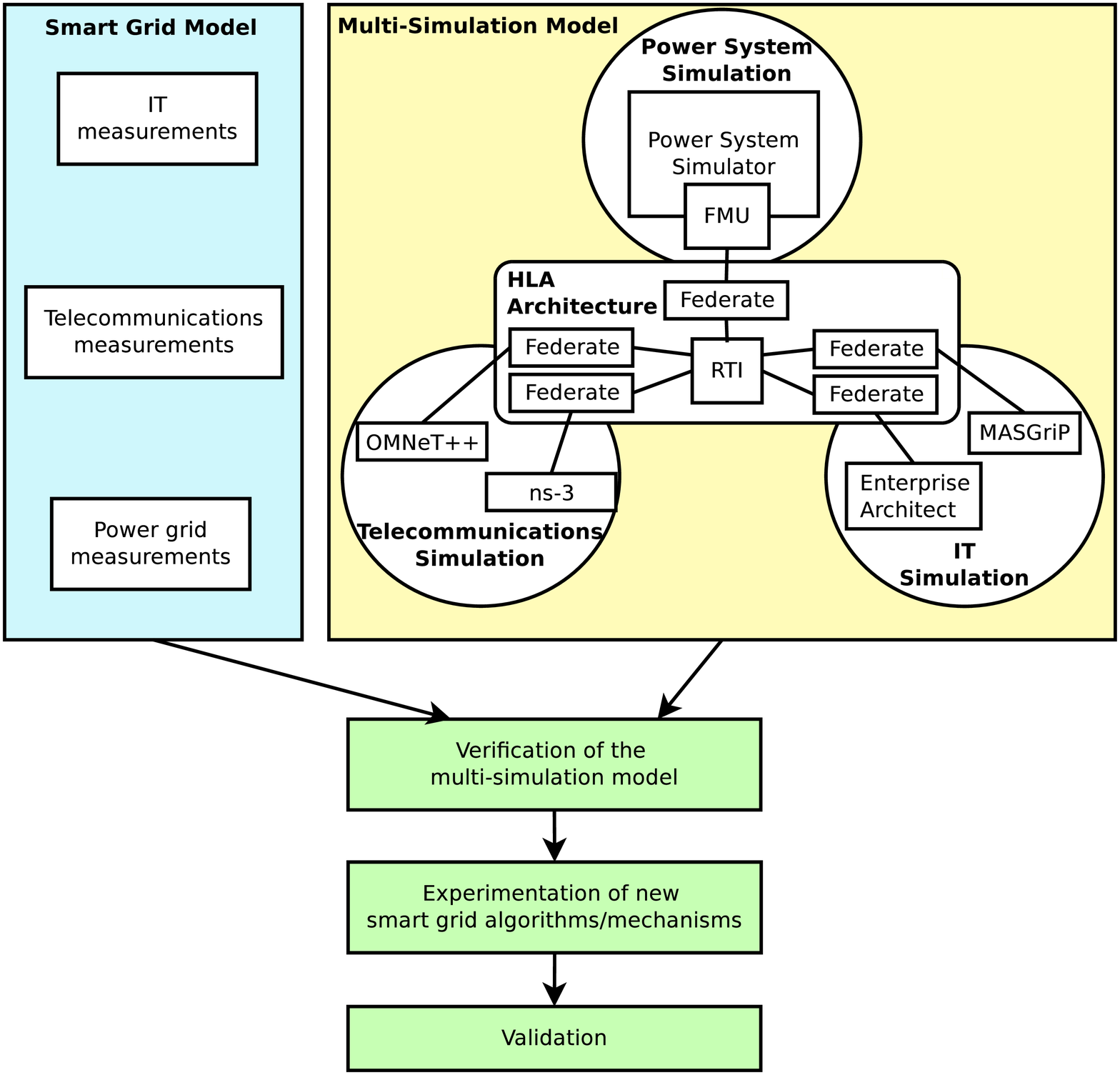}
\caption{Multi-simulation model for end-to-end validation of smart grids. Real-world measurements are compared to the obtained simulation results. The simulation model can therefore be improved to get results close to a real-world smart grid.}
\label{fig:futureImplentation}
\end{figure}

	The preliminary implementation of the IT and communication perspectives represents a step forward towards complete smart grid multi-simulations. Previous work already covered the communication and power system perspectives. The co-simulation of the IT and communication perspectives presented in this paper is novel to the best of our knowledge. Fig. \ref{fig:futureImplentation} presents our long-term vision on smart grid multi-simulations. For a given distribution grid, the utility collects measurements for each perspective for given configurations and conditions. The exact same configurations and conditions are then used in the multi-simulation model. The communication perspective is modelled using one or multiple network simulators (e.g., OMNeT++ and ns-3) depending on the technologies available in each given simulator. The simplified IT simulator presented in the previous subsection will be replaced with Enterprise Architect in future work to model complex systems using flow diagrams, Unified Modeling Language (UML), CIM models, and so forth. Enterprise Architect covers all aspects of the application development cycle, including the requirements of management, different phases of design, construction, testing, and maintenance. It also provides a complete simulator to model state machines, interactions, and activities. Note that the multi-agent simulation MASGriP, introduced in Section I, could also be used. One or multiple FMI-based power grid simulators interface with the HLA architecture.
	Results from both the smart grid and multi-simulation models are compared to verify the multi-simulator. In the case of mismatching results, the multi-simulator is improved until results get sufficiently close (e.g., less than 1\%) to real-world measurements. Once successfully verified, the multi-simulator can be used to experiment and validate new smart grid algorithms, mechanisms, and protocols, as well as test smart grid upgrades by means of multi-simulation prior to their deployment.

\section{Case Study}
\label{sec:caseStudy}

	In the following, a case study of telecontrol of DERs and distribution networks is investigated by modeling the IT and communication perspectives. As opposed to most previous studies, note that in our case study the parameter settings used for the configuration of the application, power grid topology settings, telecommunication architecture, message length, and distribution network are all based on real-world configurations, as described in greater detail in the following.

\subsection{Monitoring and Telecontrol of DERs and Distribution Networks}

\begin{figure}
\centering
\includegraphics[width=.70\textwidth]{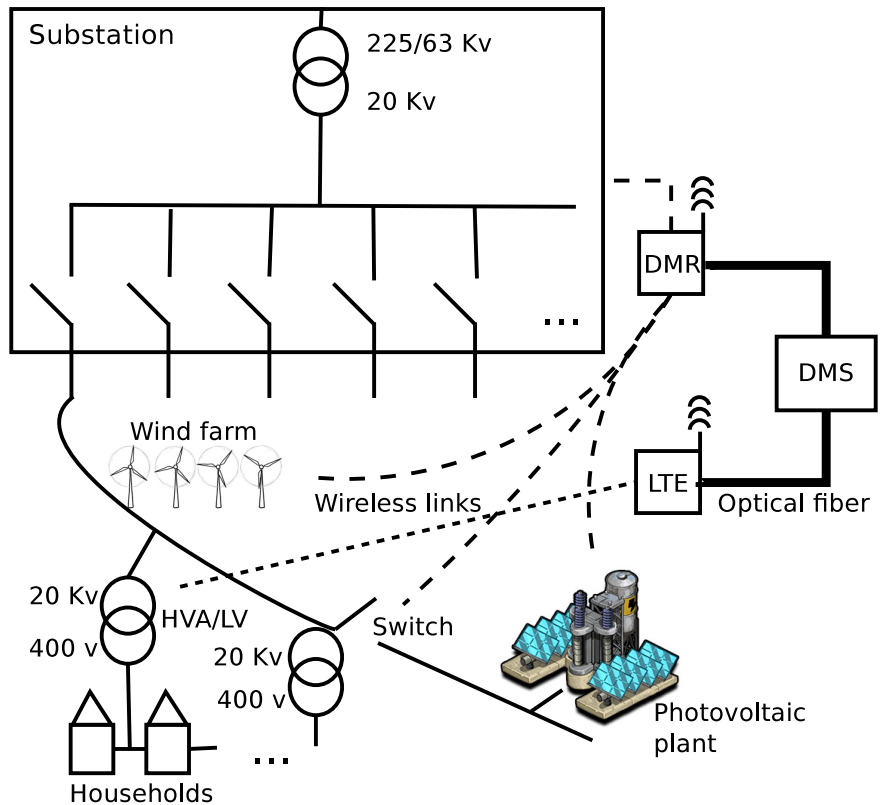}
\caption{Structure of the power distribution network interconnected via LTE and DMR wireless links. }
\label{fig:system}
\end{figure}

	The application under study is the monitoring and control of a distribution grid, as depicted in Fig. \ref{fig:system}. Monitored nodes are the following:

\begin{itemize}
\item High-voltage (20 kV)/low-voltage (400 V) nodes. 
\item Substation converting voltage from 225/63 kV to 20 kV.
\item Upcoming photovoltaic plants and wind farms.
\end{itemize}

	All monitored nodes send metrics to the distribution management system (DMS) in order to monitor the status of the distribution grid in real-time. The metrics being monitored are: active/reactive power, voltage, current, and position.

	Based on the received measurements, control messages are sent in two different scenarios:
	
\begin{itemize}
\item Automated or manual control commands are sent to the switch (ON/OFF) nodes in order to disable or activate parts of the distribution grid. Automated commands are sent to reconfigure parts of the network. Manual commands are sent before and after physical maintenance work and to manually optimize the grid.
\item As the DERs can cause the voltage profile to be outside the permissible limits ($\pm 5$ \% p.u.), the DMS sends control commands of active/reactive power in order to keep the voltage profile inside the permissible limits. 
\end{itemize}
	
\begin{figure}
\centering
\includegraphics[width=.70\textwidth]{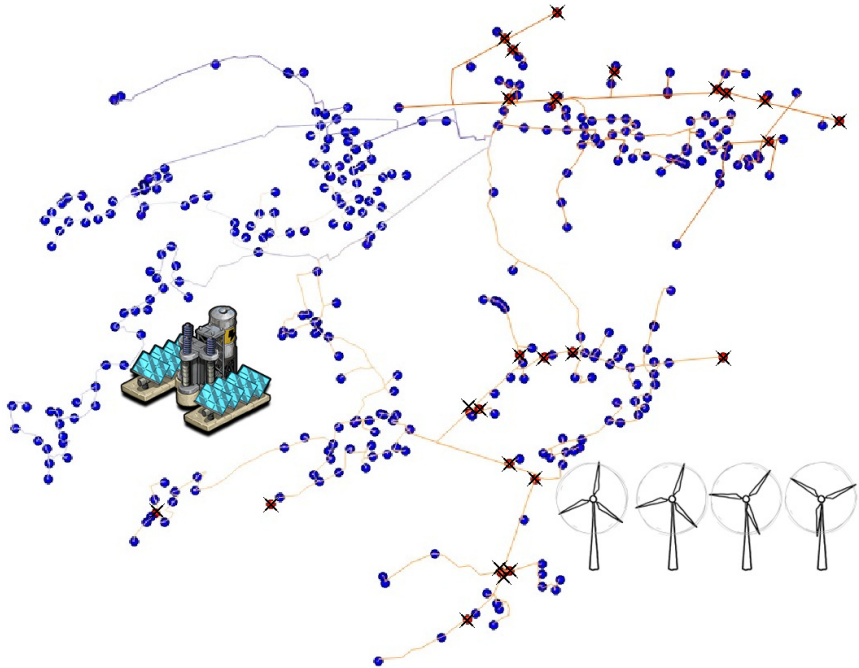}
\caption{Power distribution network topology based on real configurations in France for one given substation. Dimension: 15x15 km$^2$, number of low voltage clients: 3139, number of HVA/LV nodes (plain nodes): 332, number of switch nodes: 26 (nodes with cross). }
\label{fig:powerSysTopology}
\end{figure}

	The communication requirements of several potential smart grid applications have been determined in terms of security, reliability, bandwidth, and latency\mbox{\cite{surveySGCom}}. The authors concluded that a reliable and fast communication infrastructure is required for robust real-time exchange. In the case study considered in this paper, control commands must be delivered in high priority within 10 seconds. Monitoring messages, however, have low priority with a maximum delay in the order of 30/60 seconds. The considered power distribution network, depicted in Fig. \ref{fig:powerSysTopology}, is a real-world distribution grid in France. It has been anonymized for privacy issues, while keeping distance and position proportions. An upcoming photovoltaic plant is also added to the left-hand side of Fig. \ref{fig:powerSysTopology} and a wind farm to the bottom-right area of Fig. \ref{fig:powerSysTopology}. Note that these DER nodes are part of the simulation scenarios since we are interested in investigating their expected impact on the grid operation. DER nodes exchange both monitoring and control messages.

\subsection{Considered Communication Infrastructure and Data Flows}
\label{sec:consideredComm}
	In this work, a communication network based on Long Term Evolution (LTE) and a dedicated Digital Mobile Radio (DMR) is assumed \footnote{The DMR standard is available at \url{http://www.etsi.org/technologies-clusters/technologies/digital-mobile-radio}.}. The LTE network is used to monitor nodes and DMR to send control messages to the switches. The LTE backhaul is shared by mobile users and the utility. Therefore, at each base station a bandwidth of 50 Kbps is assumed to be reserved by the utility. The dimension of the considered topology (Fig. \ref{fig:powerSysTopology}) is 15x15 km$^2$. Furthermore, 2 LTE base stations are assumed to be available, thus totalling 100 Kbps to monitor the whole distribution network. A DMR access point is assumed to be located in the middle of the distribution network. The DMR has a low capacity of 1.920 Kbps for control messages, but it has a significant advantage to be dedicated to the utility and is thus highly reliable. Since the LTE network is not under the control of the utility the service might become unavailable which would require to re-route traffic from the LTE to the DMR network (see Section \ref{sec:impactLTEFailure}).

	The number of messages sent per second ($\lambda_m$ for monitored nodes and $\lambda_c$ for switch nodes) and average payload lengths ($\bar{L}_{dms}$, $\bar{L}_{hva/lv}$, $\bar{L}_{substation}$, and $\bar{L}_{der}$) are listed in Table \ref{table:messageConfigurations}. These configurations are used by the IT simulator to generate messages. The average payload lengths were found by capturing packets of existing applications compliant with the IEC 61850 standard. $\bar{L}_{hva/lv}$ corresponds to the average payload length of 5 manufacturing message specification messages (MMSs) and 50 MMSs for $\bar{L}_{substation}$. As for the parameter $\lambda_c$, sending 2 control commands to switches every 10 minutes might seem unrealistic (too much), it however represents a negligible percentage in terms of utilization from a communication perspective. Furthermore, as the number of DERs increase, the number of control commands will become significant. Note that 2 control commands are sent since when a problem occurs, it is often required to control 2 switches (e.g., if one link is closed, a backup link is switched ON). Note that as a preliminary implementation, only the Application and Data layer (see Fig. \ref{fig:multiSimModel}) is modeled for the IT perspective. 

\begin{table}
\caption{Experimental measurement based configurations: Messages with average payload length based on the IEC 61850 standard.}
\label{table:messageConfigurations}
\begin{center}
{\renewcommand{\arraystretch}{1.2}
    \begin{tabular}{ | l | l |}
    \hline
    Variable & Value(s) \\ \hline
    $\lambda_{m}$ & $\frac{1}{60}$, $\frac{1}{30}$  \\\hline
    $\lambda_{c}$ & $2 \cdot \frac{1}{60 \cdot 10}$ \\\hline
    $\bar{L}_{dms}$ & 64/184 bytes \\\hline
	$\bar{L}_{hva/lv}$ & 500 bytes \\\hline
	$\bar{L}_{substation}$ & 5000 bytes \\\hline
	$\bar{L}_{der}$ & 224 bytes \\\hline
	$\bar{L}_{switch}$ & 100 bytes \\\hline
    \end{tabular}
}
\end{center}
\end{table}

\subsection{IT and Communication Metrics}

	In this paper, the IT and communication perspectives are studied by considering the following metrics:
	
\paragraph{System Reliability (IT perspective)} The IT simulator is a complex information technology system, which generates messages at certain time intervals to monitor and control the power grid. It has several quality attributes and requirements, including system reliability, availability, maintainability, and so forth. Reliability is the ability to perform a certain task given a certain number of conditions and time \cite{standardP2030}. In order to quantify the reliability of the IT systems, the delay limit of the monitoring and control messages to $\mathcal{L}_\omega$ (in seconds), whereby $\omega \in \{m, c\}$ and $m, c$ correspond to the monitoring and control packet classes, respectively. Note that, unless it is explicitly related to the communication perspective, it is quantified in the IT model since the IT model generally does not have the details of the communication network, but instead simply measures the reliability metric from its perspective. For a given simulation time interval $i$, the reliability at a given node $n$ is given as follows: 

\begin{equation}
\mathcal{R}^{\omega}_{i, n} = \frac{1}{|\mathcal{M}^{\omega}_{i, n}|} \cdot \sum_{m \in \mathcal{M}^{\omega}_{i, n}} 
\begin{cases}
    1,& \text{if } d_{m, it} \leq \mathcal{L}_{\omega}\\
    0,              & \text{otherwise}
\end{cases},
\end{equation}
where $\mathcal{M}^{\omega}_{i, n}$ denotes the set of messages exchanged between $n$ and the DMS, and $d_{m, it}$ corresponds to the delay of the message $m$ from the IT perspective. Note that the equation calculates the ratio of the messages having a delay lower to the delay limit.

	For each time interval $i$, the reliability of class $\omega$ is defined according to a 95\% confidence interval:
	
\begin{equation}
\label{eq:reliability}
\mathcal{R}^{\omega}_{i} = \mu(\Theta^{\omega}_i) \pm 1.96 \cdot \frac{\sigma(\Theta^{\omega}_i)}{\sqrt{|\mathcal{N}_\omega|}},
\end{equation}
where $\mathcal{N}_\omega$ is the set of nodes exchanging messages of class $\omega$ and $\Theta^{\omega}_i$ is the distribution of the reliability at all nodes:
\begin{equation}
\Theta^{\omega}_i = [\mathcal{R}^{\omega}_{i, n}], \forall n \in \mathcal{N}_\omega.
\end{equation}

	Note that $\mu(\Theta^{\omega}_i)$ and $\sigma(\Theta^{\omega}_i)$ correspond to the mean and standard deviation of the elements in vector $\Theta^{\omega}_i$. We assume a normal distribution given that reliability is defined based on communication events, i.e., we assume that the interarrival time between messages follow a normal distribution.
 
\paragraph{End-to-end delay} From the communication perspective, the end-to-end delays corresponding to the time elapsed between the TCP senders and receivers are recorded, thus including the delays of the IP, link, and physical layers.
	
\section{Numerical Results}
\label{sec:numericalResults}

	The presented simulation case study presents different scenarios under normal conditions, with failure, with different communication QoS mechanisms, whereby the IT and communication perspectives are studied independently. 

\subsection{Influence of Simulation Timeslot Resolution}

\label{subsec:timeslot}

\begin{figure}
\centering
\includegraphics[width=0.5\textwidth]{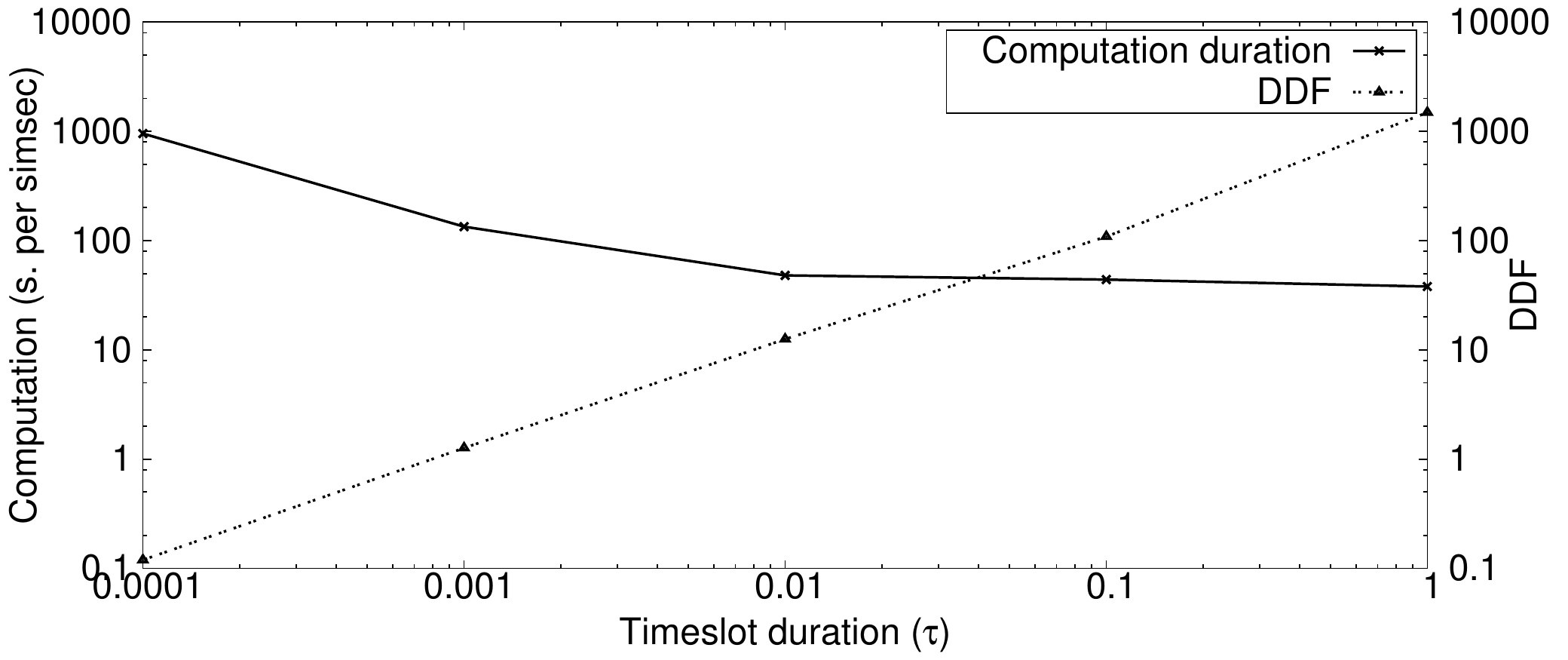}
\caption{Tradeoff between the required computation duration and DDF as a function of the timeslot duration (simsec: simulation second).}
\label{fig:computationTau}
\end{figure}

	The $\tau$ should be set significantly low to avoid synchronization errors, as outlined in previous work. However, decreasing $\tau$ increases the computation complexity since a larger number of events are created in the multi-simulator. The difference delay factor (DDF) is defined in order to define the degree of incorrectness (in percentage) of the IT perspective with respect to the communication perspective as follows:
	
\begin{equation}
DDF = 100 \cdot \frac{1}{|\mathcal{M}|} \cdot \sum_{m \in \mathcal{M}} \frac{d_{m, it} - d_{m, comm}}{d_{m, comm}},
\end{equation}
where $d_{m, it}$ and $d_{m, comm}$ account for the end-to-end delay measured in the IT and communication perspectives, respectively. Note that a perfect DDF corresponds to 0 \%, meaning that there is no delay mismatch between the IT and communication perspectives.
	Fig. \ref{fig:computationTau} illustrates the tradeoff between the DDF and required computation duration. Selecting a large value of $\tau$ leads to a high DDF, but can be computed quickly. On the other hand, selecting a low $\tau$ value leads to a lower DDF, but requires a significant amount of computation time. Note that a significant increase of the computation duration is observed, whereby 1000 seconds of real time is required to simulate 1 single simulation second. This is quite problematic from a scaling perspective. On top of this, the computation duration grows linearly as the number of nodes increases. Due to space limitations, we do not include these results. 
	

\subsection{Impact of LTE Network Failure}
\label{sec:impactLTEFailure}
\begin{figure}
\centering
\includegraphics[width=0.7\textwidth]{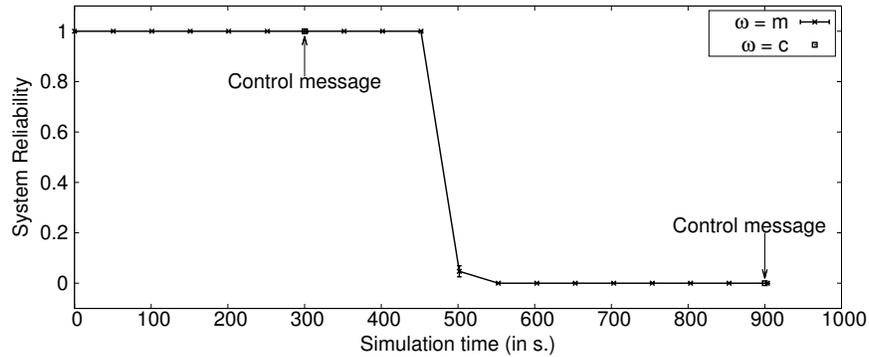}
\caption{Performance of the IT perspective for a global failure of the LTE network without using a QoS mechanism.}
\label{fig:failureResultsReliability}
\end{figure}

	As the LTE network is shared and not dedicated to the utility, the impact of a global failure of the LTE network is investigated. Monitored nodes are assumed to be equipped with an LTE and DMR interfaces. Thus, when the LTE network fails, the monitoring traffic is routed through the low capacity DMR channel. For the remainder of this section, the LTE network is assumed to become completely unavailable at simulated time 500s for ease of illustration. Fig. \ref{fig:failureResultsReliability} depicts the performance from the IT perspective. As expected, from simulation time 0 to 500, the smart grid is operated under proper conditions with low delay and high reliability. Recall from Section \ref{sec:consideredComm} that the capacity of the DMR and LTE network equals 1.92 and 50 Kbps, respectively. From simulation time 500 onwards, since all traffic is routed through the DMR network, the system reliability drops to 0 since the DMR capacity is lower than the input traffic. This is quite problematic especially for the messages controlling switches, which could cause physical damage, if not operated properly. And even from an economical perspective, not controlling switches efficiently can increase the system average interruption duration index (SAIDI) and thus increase costs to the utility.

\subsection{Improving Performance using QoS Mechanisms}

	To avoid high communication delays and keep the system reliability close to 1 when the LTE network becomes unavailable, one needs to combine both monitoring and control messages using a quality-of-service (QoS) mechanism. Note that DMR does not provide any QoS feature. In the following, the following QoS mechanisms are investigated:

\begin{itemize}
\item Weighted Fair Queueing (WFQ): WFQ is a well-known QoS mechanism, whereby each class is assigned a separate queue characterized by a weight and bandwidth is allocated according to the weight. In our work, a queue with weight 0.1 is used for monitoring messages and another one with weight 0.9 for control messages.
\item WFQ with rate adaptation (WFQ+RA): In this approach, a WFQ at the DMR access point is used, and adapt the rate at each monitored node in order to keep the network under stable conditions. When all traffic is routed to the DMR network, the data rate of the monitored nodes is updated to the following rate according to the low DMR capacity:

\begin{equation}
\lambda_m = \frac{1}{\mathcal{N}_{m}} \cdot \frac{\mathcal{T}_{total}}{(1 - \alpha_{e}) \cdot c_{dmr}},
\end{equation}
where $c_{dmr}$ denotes the DMR capacity in bits per second (bps), $\alpha_{e}$ a ratio to let some extra free capacity for retransmissions ($\alpha_{e} = 0.3$ for the remainder). $\mathcal{T}_{total}$, is the maximum amount of traffic (in bps) to transmit, including TCP/IP headers and acknowledgement packets.

The RA mechanism can be easily implemented at the application layer of the DMR communication interface without requiring any driver modification.
\end{itemize}

\begin{figure}
\centering
\begin{subfigure}[b]{0.70\textwidth}
\includegraphics[width=1\textwidth]{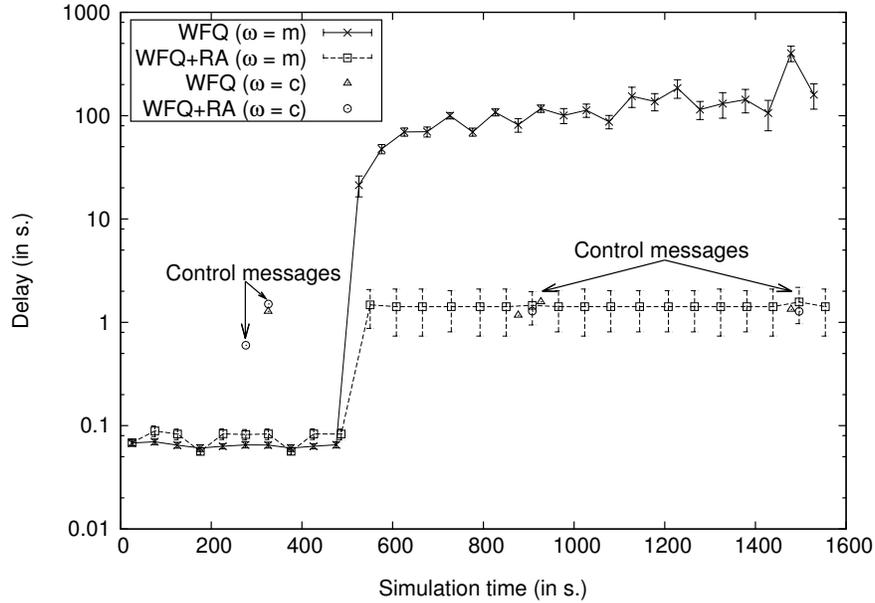}
\caption{Communication perspective.}
\label{fig:failureQoSResultsDelay}
\end{subfigure}
\begin{subfigure}[b]{0.50\textwidth}
\includegraphics[width=1\textwidth]{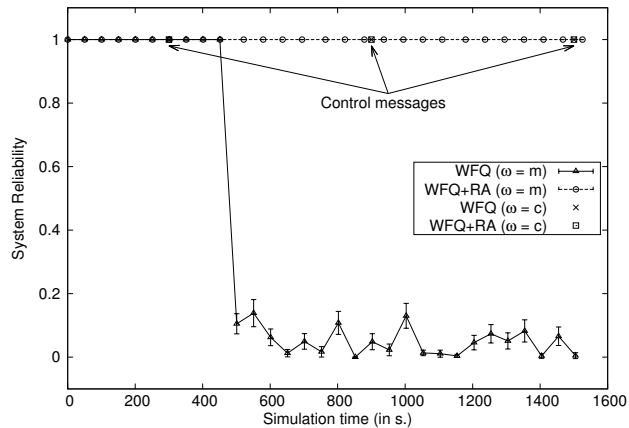}
\caption{IT perspective.}
\label{fig:failureQoSResultsReliability}
\end{subfigure}
\caption{Performance of the IT and communication perspectives in a scenario of global failure of the LTE network using different QoS mechanisms.}
\label{fig:qosFig}
\end{figure}

Fig \ref{fig:failureQoSResultsDelay} shows the communication performance in terms of delay for both QoS mechanisms. For both mechanisms, a delay close to 1 second is observed for the control messages since the WFQ processes the control messages first. However, the monitoring messages, from simulation time 500 onwards, experience high delays by using the WFQ approach since too much traffic is routed to a low-capacity link. Even worse, most traffic reaching the destination was found to correspond to the request messages from the DMS to the monitored nodes, and not the responses to the DMS. Due to space limitations, the throughput results of the response messages are not presented. The WFQ+RA mechanism experiences a low delay during the whole experiment. From simulation time 500 to 1600, the delay increases since the monitoring messages are transmitted to the DMR network. Fig. \ref{fig:failureQoSResultsReliability} shows the results from the IT perspective. Again, the reliability of the control messages remain equal to 1 because of the WFQs. However, the monitoring messages have a low system reliability from simulation time 500 onwards with the WFQ mechanism. Note that using the proposed WFQ+RA allows to keep the DMR network under stable conditions to have a system reliability close to 1.

\subsection{Validation and Improvement of the Multi-Simulation Model}
The obtained numerical results are useful to estimate potential problems stemming from routing too much traffic on a low-capacity DMR link and investigate potential solutions to avoid these problems. However, results obtained experimentally might not necessarily fully match, which would limit the usefulness of such a complex multi-simulation framework. In our future work, as outlined in Section \ref{sec:ongoingAndFutureResearch}, experimental results will be collected in order to compare the simulation results to improve the accuracy of the multi-simulation framework. Note that typically experimental results do not exactly match theoretical ones due to implementation and physical aspects which are not necessarily simulated. Therefore, to validate new smart grid applications and mechanisms by means of simulation, there is a clear need to cross-verify the multi-simulation framework using experimental results.

\section{Conclusions}
\label{sec:conclusions}
In this paper, a smart grid multi-simulation model taking into account all perspectives was presented, including the IT perspective, as standardized in IEEE P2030. Since the multi-simulation of the communication and power system has already been covered in previous work, a novel IT and communication multi-simulator was developed, whereby the power system perspective will be integrated in future work to model a complete smart grid. A case study to quantify the impacts in terms of delay and reliability when a given LTE backhaul fails was presented and all traffic is being routed through a low-capacity DMR network. To have realistic conditions in the communication perspective, real-world configurations were used for the distribution network topology and data flows based on IEC 61850 were injected. The IT and communication multi-simulator were used in order to observe problems on both perspectives, when the LTE network fails and investigated new QoS mechanisms in order to route efficiently all traffic over a low capacity DMR link. Note that the presented work in this paper represents the beginning on the modeling of complete smart grid multi-simulations, and thus unknown issues will need to be discovered. There are remaining challenges regarding multi-simulation, to name a few: ($i$) Integrating power system simulators with the HLA-based IT and communication multi-simulator might cause synchronization and performance issues and ($ii$) the obtained multi-simulator and real-world system results might mismatch, requiring to further improve the multi-simulator with respect to real-world results. The multi-simulation of smart grids is a tool to investigate new smart grid mechanisms and verify in an automated manner their impacts on all smart grid perspectives.

\bibliographystyle{IEEEtran}
\bibliography{multiSimulation}

\begin{thebibliography}{10}
\providecommand{\url}[1]{#1}
\csname url@samestyle\endcsname
\providecommand{\newblock}{\relax}
\providecommand{\bibinfo}[2]{#2}
\providecommand{\BIBentrySTDinterwordspacing}{\spaceskip=0pt\relax}
\providecommand{\BIBentryALTinterwordstretchfactor}{4}
\providecommand{\BIBentryALTinterwordspacing}{\spaceskip=\fontdimen2\font plus
\BIBentryALTinterwordstretchfactor\fontdimen3\font minus
  \fontdimen4\font\relax}
\providecommand{\BIBforeignlanguage}[2]{{%
\expandafter\ifx\csname l@#1\endcsname\relax
\typeout{** WARNING: IEEEtran.bst: No hyphenation pattern has been}%
\typeout{** loaded for the language `#1'. Using the pattern for}%
\typeout{** the default language instead.}%
\else
\language=\csname l@#1\endcsname
\fi
#2}}
\providecommand{\BIBdecl}{\relax}
\BIBdecl

\bibitem{standardP2030}
{IEEE P2030}, ``{Guide for Smart Grid Interoperability of Energy Technology and
  Information Technology Operation with the Electric Power System (EPS), and
  End-Use Applications and Loads},'' \emph{IEEE Standards Association}, Sep.
  2011.

\bibitem{geco}
{H. Lin, S. S. Veda, S. S. Shukla, L. Mili, and J. Thorp}, ``{GECO: Global
  Event-Driven Co-Simulation Framework for Interconnected Power System and
  Communication Network},'' \emph{IEEE Transactions on Smart Grid}, vol.~3,
  no.~3, pp. 1444--1456, 2012.

\bibitem{epochs}
{K. Hopkinson, X. Wang, R. Giovanini, J. Thorp, K. Birman, and D. Coury},
  ``{EPOCHS: A Platform for Agent-Based Electric Power and Communication
  Simulation Built from Commercial Off-the-Shelf Components},'' \emph{IEEE
  Transactions on Power Systems}, vol.~21, no.~2, pp. 548--558, May 2006.

\bibitem{intVGR}
{D. Q. Xu, G. Jo\'os, M. L\'evesque, and M. Maier}, ``{Integrated V2G, G2V, and
  Renewable Energy Sources Coordination Over a Converged Fiber-Wireless
  Broadband Access Network},'' \emph{IEEE Transactions on Smart Grid}, vol.~4,
  Sep. 2013.

\bibitem{coSimSurvey}
{K. Mets, J. Aparicio Ojea, and C. Develder}, ``{Combining Power and
  Communication Network Simulation for Cost-Effective Smart Grid Analysis},''
  \emph{IEEE Communications Surveys \& Tutorials (Early Access)}, Mar. 2014.

\bibitem{multiAgent2}
{P. Oliveira, T. Pinto, H. Morais, and Z. Vale}, ``{MASGriP - A Multi-Agent
  Smart Grid Simulation Platform},'' in \emph{Proc., IEEE Power and Energy
  Society General Meeting}, {San Diego, CA, USA}, Jul. 2012.

\bibitem{LTE1}
{P. Cheng, L. Wang, B. Zhen, and S. Wang}, ``{Feasibility Study of Applying LTE
  to Smart Grid},'' in \emph{Proc., IEEE First International Workshop on Smart
  Grid Modeling and Simulation (SGMS)}, {Brussels, Belgium}, Oct. 2011.

\bibitem{LTE2}
{J. Markkula and J. Haapola}, ``{Impact of Smart Grid Traffic Peak Loads on
  Shared LTE Network Performance},'' in \emph{Proc., IEEE International
  Conference on Communications (ICC)}, {Budapest, Hungary}, Jun. 2013.

\bibitem{multiAgent1}
{M. Pipattanasomporn, H. Feroze, and S. Rahman}, ``{Multi-Agent Systems in a
  Distributed Smart Grid: Design and Implementation},'' in \emph{Proc.,
  IEEE/PES Power Systems Conference and Exposition}, {Seattle, WA, USA}, Mar.
  2009.

\bibitem{coSim4}
{M. L\'evesque, D. Q. Xu, G. Jo\'os, and M. Maier}, ``{Communications and Power
  Distribution Network Co-Simulation for Multidisciplinary Smart Grid
  Experimentations},'' in \emph{Proc., SCS/ACM Spring Simulation
  Multi-Conference}, {Orlando, FL, USA}, Mar. 2012.

\bibitem{WAMPaper}
{Y. Deng, H. Lin, A. G. Phadke, S. Shukla, J. S. Thorp, and L. Mili},
  ``{Communication Network Modeling and Simulation for Wide Area Measurement
  Applications},'' in \emph{Proc., IEEE PES Innovative Smart Grid Technologies
  (ISGT)}, {Washington, DC, USA}, Jan. 2012.

\bibitem{coSim5}
{M. L\'evesque, D. Q. Xu, G. Jo\'os, and M. Maier}, ``{Co-Simulation of PEV
  Coordination Schemes Over a FiWi Smart Grid Communications Infrastructure},''
  in \emph{Proc., IEEE Industrial Electronics Society IECON}, {Montr\'eal, QC,
  Canada}, Oct. 2012.

\bibitem{HLAFirstPaper}
{R. M. Fujimoto and R. M. Weatherly}, ``{Time Management in the DoD High Level
  Architecture},'' in \emph{Proc., Workshop on Parallel and Distributed
  Simulation}, {Philadelphia, PA, USA}, May 1996.

\bibitem{fmiRTI}
{V. Liberatore and A. Al-Hammouri}, ``{The High Level Architecture RTI As a
  Master to the Functional Mock-Up Interface Components},'' in \emph{Proc.,
  International Conference on Computing, Networking and Communications (ICNC)},
  {San Diego, CA, USA}, Jan. 2013.

\bibitem{smartGridComAndCoSim}
------, ``{Smart Grid Communication and Co-Simulation},'' in \emph{Proc., IEEE
  Energytech}, {Cleveland, OH, USA}, May 2011.

\bibitem{interfacingIssues}
{X. Wang, P. Zhang, Z. Wang, V. Dinavahi, G. Chang, J. A. Martinez, A. Davoudi,
  A. Mehrizi-Sani, and S. Abhyankar}, ``{Interfacing Issues in Multiagent
  Simulation for Smart Grid Applications},'' \emph{IEEE Transactions on Power
  Delivery}, vol.~28, no.~3, pp. 1918--1927, Jul. 2013.

\bibitem{hlaOmnet}
{E. Galli, G. Cavarretta, and S. Tucci}, ``{HLA-OMNET++: An HLA Compliant
  Network Simulator},'' in \emph{Proc., 12th IEEE/ACM International Symposium
  on Distributed Simulation and Real-Time Applications}, {Vancouver, BC,
  Canada}, Oct. 2008.

\bibitem{hla2}
{H. Georg, C. Wietfeld, S. C. M\"uller, and Christian Rehtanz}, ``{A HLA Based
  Simulator Architecture for Co-Simulating ICT Based Power System Control and
  Protection Systems},'' in \emph{Proc., IEEE Third International Conference on
  Smart Grid Communications (SmartGridComm)}, {Tainan, Ta\"iwan}, Nov. 2012.

\bibitem{coSimDistrControls}
{C.-H. Yang, G. Zhabelova, C.-W. Yang, and V. Vyatkin}, ``{Co-Simulation
  Environment for Event-Driven Distributed Controls of Smart Grid},''
  \emph{IEEE Transactions on Industrial Informatics}, vol.~9, no.~3, pp.
  1423--1435, Aug. 2013.

\bibitem{coSim}
{V. Liberatore and A. Al-Hammouri}, ``{Smart Grid Communication and
  Co-Simulation},'' in \emph{Proc., IEEE Energytech}, {Cleveland, OH, USA}, May
  2011.

\bibitem{coSim2}
{H. Lin, S. Sambamoorthy, S. Shukla, J. Thorp, and L. Mili}, ``{Power System
  and Communication Network Co-Simulation for Smart Grid Applications},'' in
  \emph{Proc., IEEE PES Innovative Smart Grid Technologies (ISGT)}, {Anaheim,
  CA, USA}, Jan. 2011.

\bibitem{coSim3}
{T. Godfrey, S. Mullen, D. W. Griffith, and N. Golmie}, ``{Modeling Smart Grid
  Applications with Co-Simulation},'' in \emph{Proc., IEEE International
  Conference on Smart Grid Communications (SmartGridComm)}, {Gaithersburg, MD,
  USA}, Oct. 2010.

\bibitem{sync2}
{D. Yun, S. Kim, and S. Ha}, ``{Relaxed Synchronization Technique for
  Speeding-up the Parallel Simulation of Multiprocessor Systems},'' in
  \emph{Proc., IEEE 17th Asia and South Pacific Design Automation Conference
  (ASP-DAC)}, {Sydney, Australia}, Jan.-Feb. 2012.

\bibitem{sync3}
{D. Kim, S. Ha, and R. Gupta}, ``{Parallel Co-simulation Using Virtual
  Synchronization with Redundant Host Execution},'' in \emph{Proc., IEEE
  Design, Automation and Test in Europe (DATE)}, {Munich, Germany}, Mar. 2006.

\bibitem{sync4}
{R. Mosshammer, F. Kupzog, M. Faschang, and M. Stifter}, ``{Loose Coupling
  Architecture for Co-Simulation of Heterogeneous Components},'' in
  \emph{Proc., 39th Annual Conference of the IEEE Industrial Electronics
  Society}, {Vienna, Austria}, Nov. 2013.

\bibitem{hla1}
{J. Dingel, D. Garlan, and C. Damon}, ``{Bridging the HLA: Problems and
  Solutions},'' in \emph{Proc., IEEE International Workshop on Distributed
  Simulation and Real-Time Applications}, 2002.

\bibitem{sync1}
{X. Wang, S. John Turner, M. Yoke Hean Low, and B. Ping Gan}, ``{Optimistic
  Synchronization in HLA Based Distributed Simulation},'' in \emph{Proc., IEEE
  18th Workshop on Parallel and Distributed Simulation}, {Kufstein, Austria},
  May 2004.

\bibitem{fmi1}
{T. Blochwitz, T. Neidhold, M. Otter, M. Arnold, C. Bausch, M. Monteiro, C.
  Clau\ss, S. Wolf, H. Elmqvist, H. Olsson, A. Junghanns, J. Mauss, D.
  Neumerkel, and J.-V. Peetz}, ``{The Functional Mockup Interface for Tool
  Independent Exchange of Simulation Models},'' in \emph{Proc., The 8th
  International Modelica Conference}, {Dresden, Germany}, Mar. 2011.

\bibitem{coSimFMI}
{P. Palensky, E. Widl, M. Stifter, and A. Elsheikh}, ``{Modeling Intelligent
  Energy Systems: Co-Simulation Platform for Validating Flexible-Demand EV
  Charging Management},'' \emph{IEEE Transactions on Smart Grid}, vol.~4, Dec.
  2013.

\bibitem{cyberAttackSim}
{H. Lin, Y. Deng, S. Shukla, J. Thorp, and L. Mili}, ``{Cyber security impacts
  on all-PMU state estimator - a case study on co-simulation platform GECO},''
  in \emph{Proc., IEEE Smart Grid Communications (SmartGridComm)}, {Tainan,
  Taiwan}, Nov. 2012.

\bibitem{surveySGCom}
{V. C. Gungor, D. Sahin, T. Kocak, S. Ergut, C. Buccella, C. Cecati, and G. P.
  Hancke}, ``{A Survey on Smart Grid Potential Applications and Communication
  Requirements},'' \emph{IEEE Transactions on Industrial Informatics}, vol.~9,
  no.~1, pp. 28--42, Feb. 2013.

\end{thebibliography}
  
\end{document}